# Interference of Cooper quartet Andreev bound states in a multi-terminal graphene-based Josephson junction

Ko-Fan Huang[1†], Yuval Ronen[1†], Régis Mélin[2], Denis Feinberg[2], Kenji Watanabe[3], Takashi Taniguchi[4], Philip Kim[1,5]

[1] Department of Physics, Harvard University, Cambridge, Massachusetts 02138, USA
[2] Université Grenoble – Alpes, CNRS, Grenoble INP, Institut NEEL, 38000 Grenoble, France
[3] Research Center for Functional Materials, National Institute for Materials Science, 1-1 Namiki, Tsukuba 305-0044, Japan
[4] International Center for Materials Nanoarchitectonics, National Institute for Materials Science, 1-1 Namiki, Tsukuba 305-0044, Japan
[5] John A. Paulson School of Engineering and Applied Sciences, Harvard University, Cambridge, MA 02138, USA

[†] These authors contributed equally to this work
*Correspondence and requests for materials should be addressed to P.K. (email: pkim@physics.harvard.edu).

**In a Josephson junction (JJ), Cooper pairs are transported via Andreev bound states (ABSs)[1–3] between superconductors. The ABSs in the weak link of multi-terminal (MT) JJs[4–6] can coherently hybridize two Cooper pairs among different superconducting electrodes, resulting in the Cooper quartet (CQ) involving four fermions entanglement[7–11]. The energy spectrum of these CQ-ABS can be controlled by biasing MT-JJs due to the AC Josephson effect[12]. Here, using gate tunable four-terminal graphene JJs complemented with a flux loop, we construct CQs with a tunable spectrum. The critical quartet supercurrent exhibits magneto-oscillation associated with a charge of $4e$; thereby presenting the evidence for interference between entangled CQ-ABS[13]. At a finite bias voltage, we find the DC quartet supercurrent shows non-monotonic bias dependent behavior, attributed to Landau-Zener transitions between different Floquet bands[14]. Our experimental demonstration of coherent non-equilibrium CQ-ABS sets a path for design of artificial topological materials based on MT-JJs[15–18].**

At a normal (N)-superconductor (S) boundary, current is induced via Andreev reflection (AR)[19], *i.e.*, an electron impinging on S binds to another electron near the interface, transmitting a Cooper pair into the S region while a hole is reflected. By constructing two such boundaries one creates an SNS JJ, which can be viewed as an electronic analogue of the optical Fabry-Perot (FP) interferometer. Each boundary acts as an AR mirror and in similarity to the FP, resonances are formed in the junction. In this case, these resonances are correlated electron-hole states, the so-called Andreev bound states (ABSs). Due to Cooper pair condensation, ABSs are manifested as electron-hole pairs while each ABS is degenerate in spin. Extending this coherent transport process between multiple superconducting electrodes, two or more Cooper pairs are entangled across the N-region, forming CQ-ABS (Cooper quartet for two entangled Cooper pairs)[7–11] or higher order ABS states[15–18].

In a two-terminal JJ, two superconducting phases are introduced. However, only the phase difference $\varphi$ between the two terminals is an observable quantity. Hence the spectrum is one dimensional, depending exclusively on $\varphi$. When there is a phase difference applied across the junction, by either current or magnetic flux, each populated ABS carries a supercurrent obeying the current-phase relation and the outcome is transferring Cooper pairs from one lead to another. As a barrier is introduced in the junction, a gap is opened between each electron-hole pair of equal spin states. Therefore, the energies have a $2\pi$-periodicity with respect to $\varphi$ and the ABS spectrum is analogous to a crystal band structure with $\varphi$ playing the role of crystal momentum. More generally, the ABS spectrum may span in 2D, 3D or even higher dimensions, solely determined by the number of superconducting contacts on the same junction.

Along with the ability of controlling the number of conducting channels, low superconducting contact resistance and weak back-scattering[20–22] make graphene an ideal choice for exploring MT-ABS physics. Utilizing the tunability of graphene chemical potential, one can modulate the coupling strength at each contact, thereby engineering the ABS spectrum. Our graphene-based MT-JJs use Ti/Al as the superconducting contacts, where Al is chosen owing to its large superconducting coherence length (~1μm).

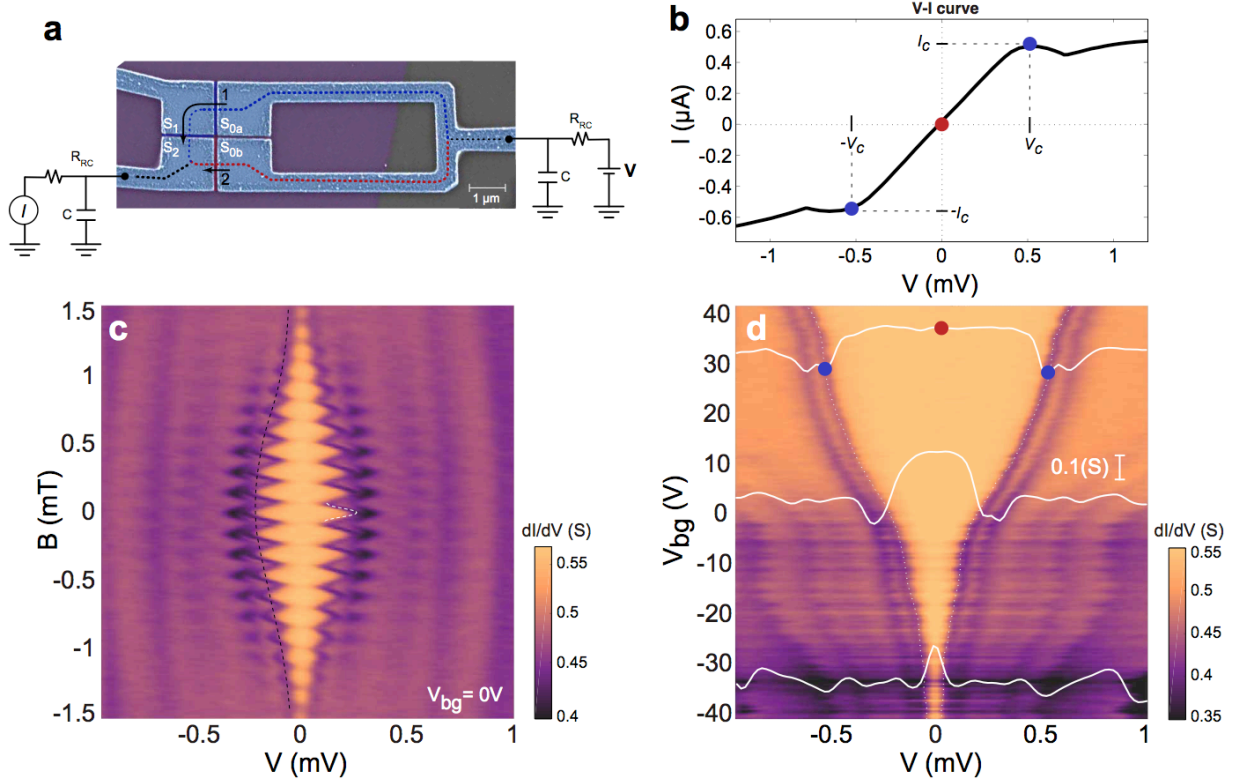

**Figure 1. Single source voltage bias characterization of four-terminal Josephson junction including a loop. a**, False color scanning electron microscopy (SEM) image of the device with measurement configurations. Graphene (purple) is top-contacted by Ti/Al superconducting electrodes (blue) and the electrode separations typically are 80-100 nm. **b**, $V - I$ curve of the device from the measurement configuration in **a**. $I_c$ is the critical current and the corresponding voltage value is labelled as $V_c$ (the blue dots). **c**, Magnetic field dependence, $dI/dV$ as a function of the bias voltage and magnetic field. Bright orange region (high conductance) is the supercurrent and the edge corresponds to the value of critical current, which is modulated by the magnetic field. The SQUID-like pattern indicates the interference between two supercurrent paths (red and blue dashed lines in **a**). The periodicity of the fast oscillation (white dashed curve) corresponds to the loop area and the slow oscillation (blacked dashed curve) is the first lobe of Fraunhofer pattern. **d**, Gate dependence of the supercurrent, $dI/dV$ as a function of the bias voltage and global back-gate voltage $V_{bg}$. The critical current reaches the minimum as graphene is tuned to the Dirac point near $V_{bg} = $ -32 V.

A four-terminal JJ including a superconducting loop is fabricated on graphene-hBN-SiO$_2$ structure as shown in Fig. 1a (additional fabrication information can be found in the Method section).

All measurements were performed at 300 mK. Before we conduct the MT-JJ measurement, we first characterize our device with a two-terminal measurement and the S-loop implements a superconducting quantum interference device (SQUID) geometry. For this measurement, we applied a bias voltage $V$ to the loop, via two series connected RC filters and the output current $I$ is measured at $S_2$ while $S_1$ is floating. Fig. 1b shows an $I$-$V$ measurement curve of the junction. In the small bias regime, supercurrent flows in the junction and the bias voltage drops are only on the series connected resistors $R_{RC}$ (200 Ω each) in the filters. As the current exceeds the critical current $I_c$ of the SQUID, the slope of $I$-$V$ curve changes suddenly at the corresponding applied voltage $V_c$. Since the bias voltage is distributed among two filter resistors and the normal junction resistance, the critical current can be obtained from $I_c = V_c/2R_{RC}$. Upon applying the magnetic field $B$, $I_c$ is modulated and exhibits SQUID-like pattern as a result of the two interfering superconducting paths in the loop (blue and red dashed lines in Fig 1a). Fig. 1c shows the differential conductance ($G = dI/dV$) as a function of bias voltage and magnetic field. The higher conductance area near the zero-bias regime (central orange part) is the supercurrent region and its edges mark the value of $I_c$. As the magnetic field is swept, $I_c$ is modulated with a periodicity of $\delta B = 145$ μT, corresponding to the unit flux quantum $\Phi_0 = h/2e$ for an enclosed area of $A = 14.2$ μm², matching our device loop size (including the area increase due to London penetration depth). An additional slower frequency ($\delta B_F = 3$ mT) originated from the Fraunhofer oscillations is observed, corresponding to an area of 0.69 μm², which agrees with the junction dimensions. We find that the strength of the critical current can also be tuned according to the graphene carrier density via a back-gate voltage $V_{bg}$. As shown in Fig. 1d, $I_c$ decreases monotonically as $V_{bg}$ approaches the charge neutrality point of graphene located at $V_{bg} \approx -60$ V. Reduction of $I_c$ close to the Dirac point is expected due to the decreasing number of ABS carrying current in the graphene channels[23].

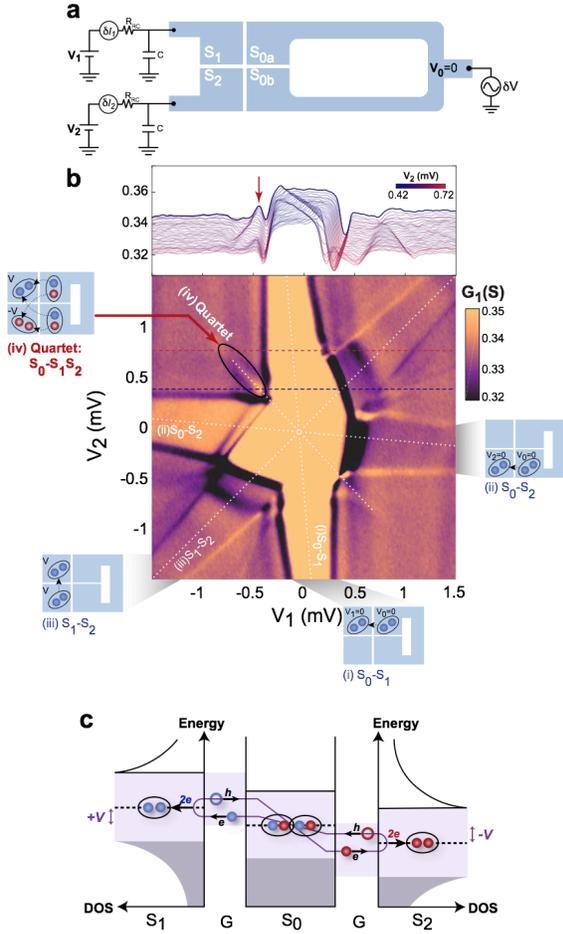

**Figure 2. Dual source voltage bias characterization for quartet detection. a**, Configuration for the quartet measurement. The loop $S_0$ is grounded while potential of the first ($S_1$) and second ($S_2$) electrodes are controlled via DC voltages $V_1$ and $V_2$, respectively. An additional AC excitation $\delta V$ is applied to the loop, and the AC currents $\delta I_1$ ($\delta I_2$) through $S_1$ ($S_2$) is measured. **b**, top panel: differential conductance $G_1$ (= $\delta I_1/\delta V$) measured at $S_1$ as a function of the DC bias voltage $V_1$ when $V_2$ is tuned from 0.42 mV to 0.72 mV. Bottom panel: color plot of $G_1$ as a function of $V_1$ and $V_2$. In total, there are four different supercurrents in the device. Inset (i)-(iii) shows the local Josephson supercurrent between any pair of leads at the same potential. Inset (iv) Quartet shows the nonlocal quartet supercurrent flowing among all three superconducting leads and the quartet signal in ($V_1$, $V_2$)-plane is the narrow yellow region along the − 45 degree direction (where the red arrow points at). **c**, Schematic illustration of the three-terminal quartet process with the Andreev reflection picture. The middle superconductor $S_0$ is grounded while the other two superconductors are biased at $+V$, $-V$, respectively. The two entangled Cooper pairs (with red and blue electrons) are formed in $S_0$ through two local Andreev reflections and two cross Andreev reflections.

With reconfiguration of the external circuitry, our device can serve as a MT-JJ where the common N-region graphene channel is proximitized. MT-JJ with magnetic flux loops was studied theoretically and experimentally in bi-SQUID devices[24,25], where the equilibrium (*i.e.*, no potential difference between the junctions) ABS spectrum was investigated. Our four-terminal device geometry with gate-tunable graphene weak link allows us to study biased MT-JJs in the non-equilibrium regime, where the nonlocal CQ can be investigated[26]. Moreover, by threading a flux through the device loop we aim to modulate the CQ-ABS spectrum. Fig. 2a shows the measurement scheme adopted in this study for phase sensitive quartet detection. We apply DC bias voltages $V_1$ and $V_2$ to $S_1$ and $S_2$, respectively, and a small AC bias voltage $\delta V$ to the loop electrodes $S_{0a}$ and $S_{0b}$. At given bias voltages, we measure the AC current contribution $\delta I_1$ and $\delta I_2$ flowing to $S_1$ and $S_2$, respectively.

Figure 2b shows the differential conductance measured at $S_1$ ($G_1 = \delta I_1/\delta V$) as a function of the two DC bias voltages $V_1$ and $V_2$. We identify four high conductance regions (marked by four white dashed lines crossing at the origin), which correspond to four different supercurrents. For instance, when $S_2$ and $S_0$ are equipotential along $V_2 = 0$, a Josephson supercurrent flows between these two contacts carried by a Cooper pair-ABS. Subfigures (i) (ii) and (iii) illustrate these local supercurrents between different pairs of S-contacts. The critical values of the supercurrents can be extracted from the widths of the signals, which are 0.47, 0.42, 0.38 μA, respectively. Similar data can be obtained for differential conductance $G_2 = \delta I_2/\delta V$ measured at $S_2$ (see Section 2 in SI).

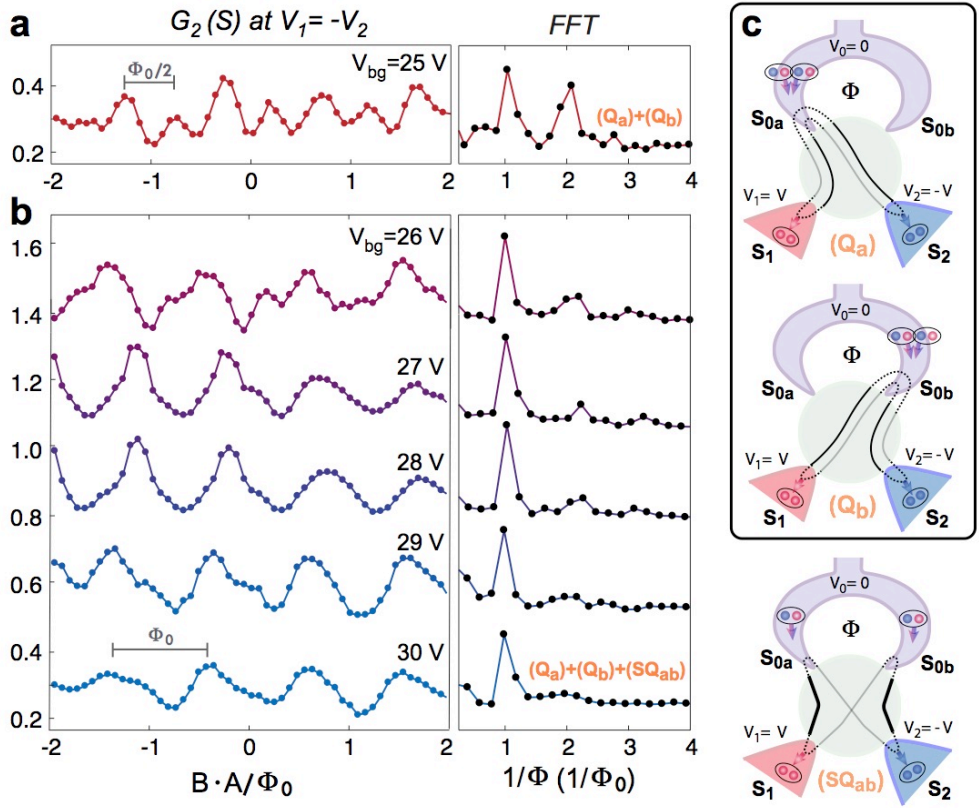

**Figure 3. Different types of quartet process. a**, Left panel shows the quartet differential conductance $G_2$ ($= \delta I_2/\delta V$) measured at $S_2$ a function of the flux $\Phi = B \cdot A/\Phi_0$ for $V_{bg} = 25$ V. Satisfying the quartet bias condition, $V_1 = -V_2$ is fixed at 4 V. Right panel shows the analysis from fast Fourier transform (FFT) with a prominent second harmonic. **b**, for $V_{bg} = 26$-30 V. The periodicity evolves from half-flux quantum to one flux quantum as $V_{bg}$ increases. **c**, Subfigure ($Q_a$)($Q_b$) shows the conventional three-terminal quartet process with only one out of the two loop contacts involved. Electron-hole conversion happens twice at the same contact of the loop (either $S_{0a}$ or $S_{0b}$), resulting in periodicity of half-flux quantum. Subfigure (SQ$_{ab}$) shows the split-quartet process involving both contacts of the loop. With the odd parity of Cooper pairs transferred, the periodicity is one flux quantum.

In addition to the two-terminal Josephson currents (i)-(iii), we observe another supercurrent signal along the $V_1 = -V_2$ line. This supercurrent is labelled as (iv) Quartet and it is carried by CQ-ABS. The quartet supercurrent is a direct evidence of MT-ABSs in our system due to the fact that all contacts are at different chemical potentials. Fig 2c describes the formation of CQ-ABS: two Cooper pairs from two S-contacts ($S_1$ and $S_2$) are entangled into a four-electron state via two local Andreev reflections and two cross Andreev reflections at the middle S-contact ($S_0$)[7–11]. Remarkably, at the high bias regime where the local DC-Josephson currents disappear, the quartet ABSs persist when the junction is biased anti-symmetrically and carry nonlocal supercurrent flowing among all terminals simultaneously. The corresponding bias condition $V_1 + V_2 = 0$ satisfies the energy conservation for the CQ-ABS, where correlated Cooper pairs originating from $S_1$ and $S_2$ are transmitted into $S_0$.

Similar quartet supercurrent signatures were inferred previously in three-terminal Josephson junctions made from diffusive metal[10] and 1D nanowires[11]. The novelty in our four-terminal JJ device is the presence of a magnetic flux loop, enabling direct probing of the CQ-ABS coherence via magnetic field dependence of the critical current. The left panels of Fig. 3a-b show the quartet differential conductance measured at $S_2$ (i.e., $G_2$ along the quartet line $V_1 = -V_2$) as a function of the magnetic flux $\Phi = B \cdot A$ measured at different back-gate voltage $V_{bg}$. The quartet differential conductance $G_{i=1,2}$ probes the quartet critical current $I_{qc}$ (see Section 5 in SI). As a function of $\Phi$, clear oscillations of $G_i$ are observed, demonstrating periodic modulation of $I_{qc}(\Phi)$ due to phase coherence of the CQ-ABS. By taking the Fourier transform of $G(\Phi)$ (the right panel of Fig. 3 a-b), we find two major periodicities $\Phi_0/2$ and $\Phi_0$, where $\Phi_0 = h/2e$. The relative strength of the periodicities is tuned via $V_{bg}$. As we discussed in Fig. 1d, $V_{bg}$ adjusts the number of channels in graphene and the coupling of S-electrodes, thus modifying the ABS spectrum. In particular, at $V_{bg} = 25$ V (Fig. 3a), $I_{qc}$ exhibits a prominent contribution from $\Phi_0/2$-periodicity, which, as we show now, provides direct evidence for the charge $4e$ associated with the CQ-ABS.

The observation of the two periodicities tuned by the gate voltage resembles the SQUID oscillation in Fig. 1c, where the $\Phi_0/2$ oscillation would be viewed as the second harmonic of the fundamental quantum flux periodicity. However, the magneto-oscillation here in Fig. 3a-b cannot be related to DC-SQUID harmonics since there is only one single junction (between $S_{0a}$ and $S_{0b}$) that is at equal potential. In addition, the bias voltages used in this measurement are in a range where one has AC Cooper pair Josephson currents rather than DC. Furthermore, we carefully delineate our signal from the contribution of finite bias multiple Andreev reflections (MARs) by measuring an oscillatory differential conductance, following the $I_{qc}(\Phi)$ variation along the quartet bias condition ($V_1 = -V_2$) above the MARs background (see Fig. S3c in SI). By adopting the perturbative approach expanded towards the finite bias regime (see Section 5 in SI for detailed discussion), we find that the modulation of the periodicity is associated with interference of three different contributions to the CQ-ABS: two conventional quartets (3-terminal) and a novel process, the split-quartet (4-terminal). As shown in Fig. 3c ($Q_a$) & ($Q_b$), the two conventional quartets take place among $S_1$, $S_2$ and only one of the two loop electrodes. In these processes, the entangled Cooper pairs enter the loop either through $S_{0a}$ or $S_{0b}$. Since every Andreev reflection picks up the phase of the superconducting contact, these conventional quartet processes acquire phase factors $e^{i(\varphi_1+\varphi_2)}$ at $S_{0a}$ and $e^{i(\varphi_1+\varphi_2+4\pi\Phi/\Phi_0)}$ at $S_{0b}$, where $\varphi_1$ ($\varphi_2$) is the phase difference between $S_1$ ($S_2$) and $S_0$. Note that the factor 4 in the exponent reflects that two Cooper pairs depart from the same electrode of the grounded loop. If there were only this type of 3-terminal quartet process in the system, the phase factor at $\Phi/\Phi_0 = 0$ would become equivalent to that at $\Phi/\Phi_0 = 1/2$, leading to $\Phi_0/2$-periodicity in $I_{qc}(\Phi)$.

While the conventional quartet process described above is common in three terminal JJs, the four-terminal JJ with a loop enables a different type of quartet, the split quartet (Fig. 3c (SQ$_{ab}$)). In the split quartet process, two entangled Cooper pairs are spatially separated into the two electrodes of the loop, yielding a phase factor $e^{i(\varphi_1+\varphi_2+2\pi\Phi/\Phi_0)}$. The resulting critical current contribution has $\Phi_0$-periodicity. The modulation of $\Phi_0/2$- and $\Phi_0$-periodic oscillation components in $I_{qc}(\Phi)$ at different back-gate voltages indicates that the occurrence of the two different types (i.e., conventional and split) of quartet process is determined by the relative contact couplings, which are tunable via gating (see Section 5 in SI).

The quartet supercurrent can also be modulated by the quartet voltage $V_q$, which is the actual voltage applied on the junction along $V_1 = -V_2$. The variation of $G(\Phi, V_q)$ is proportional to that of $I_{qc}(\Phi, V_q)$ along the quartet line since it is an increasing function of the critical current (see section 5 in SI). Therefore this differential conductance measurement serves as a good indicator to investigate the behavior of quartets as a function of magnetic field and the quartet voltage. Figure 4a shows a 2D color plot of $G_1$ (the quartet conductance measured at $S_1$) as a function of $V_q$ and the normalized magnetic flux $\Phi/\Phi_0$ at a fixed gate voltage $V_{bg} = -5$ V, where the quartet current is strong (see Fig. S4 in SI). At a constant $V_q$, $G_1(\Phi)$ exhibits oscillations corresponding to $I_{qc}(\Phi)$ with periodicity $\Phi_0/2$ and $\Phi_0$ components as discussed in Fig. 3. Interestingly, we find that the oscillation period and phase of $I_{qc}(\Phi)$ are also tunable as $V_q$ varies. As shown in Fig. 4b, in the low bias regime ($V_q < 7.2$ µV), $I_{qc}(\Phi)$ shows predominantly the $\Phi_0$-periodic oscillation, in phase with the SQUID phase of equilibrium supercurrent. However, as $V_q$ increases, $I_{qc}(\Phi)$ oscillation becomes predominantly $\Phi_0/2$-periodic near $V_q \approx V_{in} \equiv 7.2$ µV. Above this critical bias voltage $V_{in}$, $I_{qc}(\Phi)$ oscillation resumes the $\Phi_0$-period, but the phase is shifted by $\pi$ compared to $I_{qc}(\Phi, V_q < V_{in})$. We note that for this high bias quartet regime ($V_q > V_{in}$), the flux dependence of the quartet critical current is « inverted », i.e., $I_{qc}(\Phi = 0) < I_{qc}(\Phi = \Phi_0/2)$, suggesting an unusual quartet behavior occurs as we approach the high bias limit.

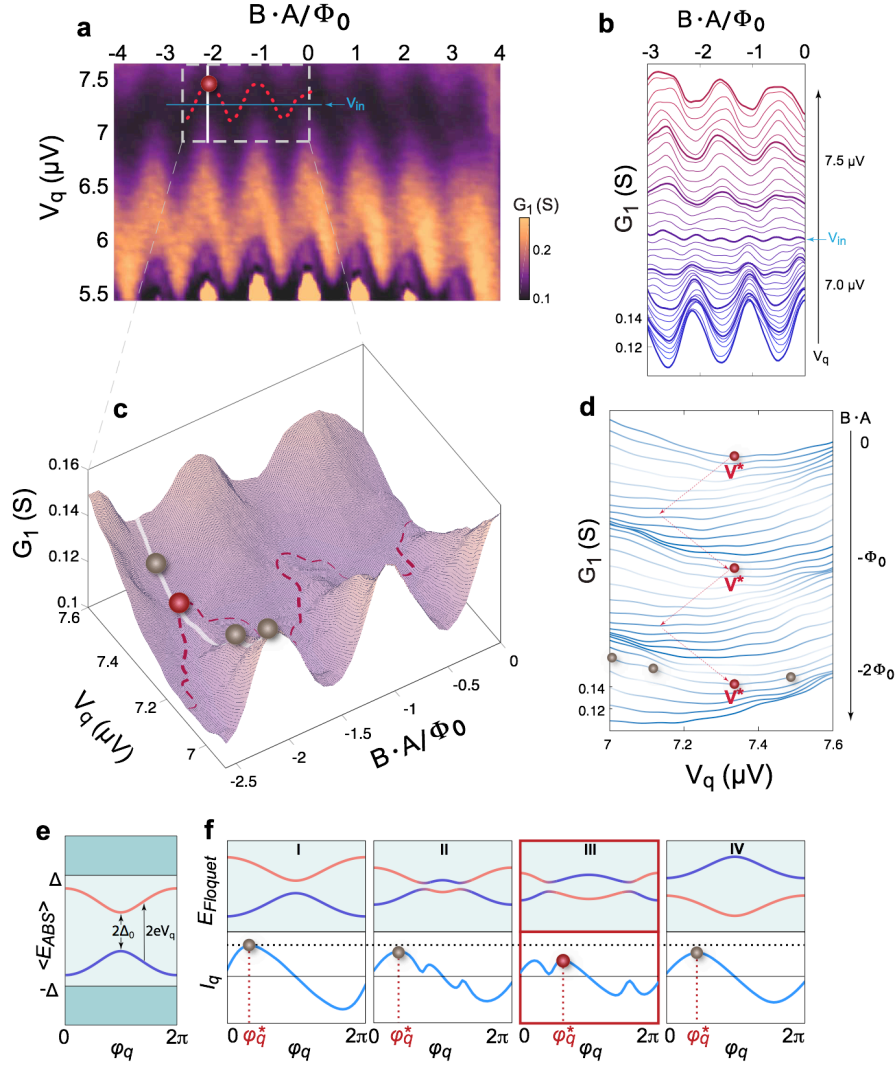

**Figure 4. Quartet conductance and the Floquet spectrum as a function of $V_q$ and magnetic flux. a**, Quartet conductance $G_1$ measured along the quartet line in $(V_q, \Phi)$-plane. The red dashed line traces the minimum conductance for $-2.5 < -\Phi/\Phi_0 < 0$ and the red sphere represents the local minimum at $\Phi/\Phi_0 = -2$. **b**, waterfall plot of $G_1(\Phi)$ for $V_q = 6.9\text{-}7.6\ \mu V$. It shows clear evolution of $G_1$ from maxima to minima at inter values of flux. At the critical quartet $V_q \approx V_{in} \equiv 7.2\ \mu V$, periodicity is $\Phi_0/2$ and for $V_q > V_{in}$, the quartet critical current is « inverted ». **c**, Zoom-in surface plot of $G_1(V_q, \Phi)$ for $-2.5 < \Phi/\Phi_0 < 0$. The winding of the red sphere (local minimum) is marked with the red dashed line, matching that in Fig 4a. The gray spheres represent quartet conductance at different values of $V_q$. **d**, waterfall plot of $G_1(V_q)$ for $\Phi/\Phi_0 = -2 \sim 0$. The local minimum $V^*$ presents a zig-zag pattern as flux is tuned. **e**, When the quartet voltage $V_q$ is in the adiabatic limit, the effective Andreev levels $<E_{ABS}>$ depend only on one phase variable, the quartet phase $\varphi_q$. The minimum difference between the two levels is the Andreev gap $\Delta_0$ and a finite $V_q$ creates resonant coupling between the two levels. **f**, upper panel shows the energy of the Floquet states as a function of the quartet phase $\varphi_q$ at different values of $V_q$. The corresponding quartet current $I_q$ carried by these Floquet states is shown in the lower panel. The gray and red spheres mark the critical values of the quartet current $I_{qc} = I_q(\varphi_q^*)$, matching the ones in Fig. 4c. In (III), the red sphere denotes $V_q = V^*$ when $I_{qc}$ reaches a local minimum, reflecting an avoided crossing in the Floquet spectrum.

Furthermore, for a fixed $\Phi$, $G_1(V_q)$ displays distinct non-monotonic behavior as $V_q$ varies near $V_{in}$. As shown in Fig. 4c (the zoom-in 3D map) and Fig. 4d (line-cuts for flux in the range $[-2.5\Phi_0, 0]$), $G_1(V_q)$ first decreases to yield a local minimum at $V_q \approx V^*$ (represented by the red sphere) and then increases for

$V_q > V^*$. We note that $V^*$ shifts in a zig-zag pattern in the $(V_q, \Phi)$-plane centered at $V_{in}$. Particularly, $V^*(\Phi)$ is the largest at integer $\Phi_0$ and the smallest at half-integer $\Phi_0$, similar to the inverted quartet current in previous discussion. The non-monotonic variation of $I_{qc}(V_q)$ and the inversion of the quartet current flux dependence provide a clue to the dynamic behavior of quartets in the non-equilibrium condition at a finite $V_q$.

To understand the $V_q$ dependence of quartet current, we need to consider the superconducting phase modulation due to the AC Josephson effect at finite bias. We first set the phase of the grounded loop $S_0$ to be zero and the other two superconducting leads $S_1$ and $S_2$ have phases $\varphi_1$ and $\varphi_2$, respectively. When voltages are applied to $S_1$ and $S_2$, the phases acquire time ($t$)-dependence following the Josephson relation[12]: $\dot\varphi_1(t) = \frac{2eV_1}{\hbar}$, $\dot\varphi_2(t) = \frac{2eV_2}{\hbar}$. Under the quartet condition ($V_1 = -V_2$) and by choosing a new set of phase variables $\varphi_q \equiv \varphi_1(t) + \varphi_2(t), \varphi_r \equiv \varphi_1(t) - \varphi_2(t) = \frac{4eV_q}{\hbar}t$, we obtain a stationary quartet phase $\varphi_q$ and a running phase $\varphi_r$ that is periodically driving the system with a frequency $4eV_q/h$. In the adiabatic limit, i.e., $V_q$ is much smaller than the Andreev minigaps $\Delta_0$ between the ABS pairs, one can take a time average of the ABS spectrum over $\varphi_r$ and obtain an effective ABS energy $<E_{ABS}>$, which now only depends on the quartet phase $\varphi_q$ (see Section 5 in SI). For simplicity, we consider only a single pair of ABSs at a small bias, where the adiabatic approximation works. $I_q$, the supercurrent carried by quartets, can then be derived from the usual JJ current-phase relation: $I_q = \frac{2e}{\hbar}\partial<E_{ABS}>/\partial\varphi_q$. However, as $V_q$ increases, we eventually enter the non-adiabatic regime, where occupation of higher ABSs must be accounted for. In this regime, the AC Josephson effect creates resonant coupling between the two adiabatic ABS levels (Fig. 4e). As a result, the effective separation of the two ABSs can be adjusted by the bias voltage in analogy to the Floquet bands[27,28] separated by $2eV_q$, emerging from the periodically driven Bloch bands[26,29].

Employing the Floquet energy levels $E_{Floquet}$ that are derived from a pair of $<E_{ABS}>$ biased by the quartet bias $V_q$ (Fig. 4e), we can now explain the experimentally observed non-monotonic behavior of $I_{qc}(V_q)$. Figure 4f shows the evolution of two first-order $E_{Floquet}$ as a function of quartet phase $\varphi_q$. The corresponding quartet current $I_q(\varphi_q)$, shown in the bottom panels, is obtained with the Floquet-Landau-Zener[30] consideration (see Section 5 in SI) and the critical quartet current $I_{qc}(= max\{I_q(\varphi_q)\})$ takes place at $\varphi_q^*$. As $V_q$ increases, four different regimes appear: (I) for $2eV_q < \Delta_0$, no resonant coupling exists between the two $<E_{ABS}>$ and the quartet current is the same as near equilibrium. (II) $2eV_q \sim \Delta_0$, i.e., the Landau-Zener (LZ) transitions between the two $<E_{ABS}>$ bands become appreciable, opening gaps between different $Floquet$ bands. Hybridization between two levels and mixing of states that carry opposite directions of currents reduce the net quartet current, resulting in a drop in $I_{qc} = max\{I_q(\varphi_q)\}$) and the shifting of $\varphi_q^*$. (III) At even larger quartet voltage $V_q = V^*$, the resonances occur at the $\varphi_q^*$ in (I), denting the peak in $I_q$ and thereby $I_{qc}$ reaches a minimum value. (IV) When $2eV_q$ is increased to be greater than the largest gap between the two levels, there is no more hybridization. Both the energy levels and the quartet current resume the nearly adiabatic situation, similar to (I). For a more accurate consideration, the non-equilibrium Keldysh formalism is applied to multi-level ABSs. It reveals that the inversion of $I_{qc}(\Phi)$ can be associated with the avoided crossings due to LZ transition in the Floquet bands (see Section 5 of SI).

**Methods** The van der Waals heterostructure – monolayer graphene on top of 40-60 nm thick hBN – is assembled via the inverted stacking technique, where hBN serves as the dielectric substrate to minimize disorder[31]. The flakes are picked up through procedure similar to the dry transfer technique[32] except the order is reversed, where the bottom hBN is picked up first. Via this method the top surface of graphene is guaranteed to be clean without any polymer contact in the assemble process. The superconducting contacts are made of 80 nm thick aluminum with 5 nm thick sticking layer of titanium, directly deposited on graphene through electron-beam evaporation at a pressure of low $10^{-7}$ torr. Each channel is designed to be 80-90 nm to ensure the existence of supercurrents among all of the superconductors. The measurements are performed in He-3 fridge with the base temperature 300 mK, well below the superconducting critical temperature of aluminum ($T_c \sim 1.1$ K) and the dual voltage source measurement scheme allows the detection of quartet signal (see Supplementary Information for the details).

**Acknowledgements** We thank B. Douçot for his collaboration on the Floquet theory. K.-F.H. acknowledges support from DOE (DE-SC0019300) for sample preparation and fabrication. Y.R. acknowledges support from NSF (QII-TAQS MPS 1936263) for device characterization. P.K. acknowledges NSF (DMR1809188) for data analysis. R.M acknowledges the use of the resources of the Mésocentre de Calcul Intensif de l'Université Grenoble-Alpes (CIMENT). K.W. and T.T. acknowledge support from the Elemental Strategy Initiative conducted by the MEXT, Japan, Grant Number JPMXP0112101001, JSPS KAKENHI Grant Number JP20H00354 and the CREST(JPMJCR15F3), JST.

**Author contributions** Y.R., K.-F.H. and P.K. designed the experiment. P.K. supervised the project. K.-F.H. and Y.R. fabricated the devices. T.T and K.W. provided single crystals of hBN. K.-F.H. and Y.R. performed the measurements. K.-F.H., Y.R., and P.K. analyzed the data. R.M. and D.F. carried out the theoretical modelling. K.-F.H., Y.R., and P.K. prepared the manuscript and Supplementary Information with input from all the authors.

SUPPLEMENTARY INFORMATION

**Interference of Cooper quartet Andreev bound states in a multi-terminal graphene-based Josephson junction**

Ko-Fan Huang[1†], Yuval Ronen[1†], Régis Mélin[2], Denis Feinberg[2], Kenji Watanabe[3], Takashi Taniguchi[4], Philip Kim[1,5] *

[1] Department of Physics, Harvard University, Cambridge, Massachusetts 02138, USA

[2] Université Grenoble – Alpes, CNRS, Grenoble INP, Institut NEEL, 38000 Grenoble, France

[3] Research Center for Functional Materials, National Institute for Materials Science, 1-1 Namiki, Tsukuba 305-0044, Japan

[4] International Center for Materials Nanoarchitectonics, National Institute for Materials Science, 1-1 Namiki, Tsukuba 305-0044, Japan

[5] John A. Paulson School of Engineering and Applied Sciences, Harvard University, Cambridge, MA 02138, USA

† These authors contributed equally to this work

* Corresponding author: pkim@physics.harvard.edu

## S1. Dual voltage source for quartet measurement

In order to control the potential of each superconducting terminal, the quartets are detected through the dual voltage source measurement scheme (Fig. S1). Terminal $S_1$ and $S_2$ are biased with DC voltages $V_{b1}$ and $V_{b2}$ through a voltage divider and an RC filter (in the main text the voltage dividers are not shown). The loop terminal $S_0$ remains grounded at all times but on top of this DC ground, we apply a small AC excitation in the range of 0.25-0.3 V. Like the other leads, this lead also has a voltage divider followed by an RC filter. The voltage divider for the loop $S_0$ divides the AC excitation by $10^5$. In order to detect the quartet current, we measure the conductance at the biased terminals $S_1$ and $S_2$. As shown in the circuit, we use lock-in amplifier to measure the potential $dV_1$ and $dV_2$. The AC currents owing from are then given by $dI_i = dV_i/r$, where $i = 1,2$. Therefore, the conductance at each lead is $G_i = dI_i/dV = dV_i/(r \cdot dV)$.

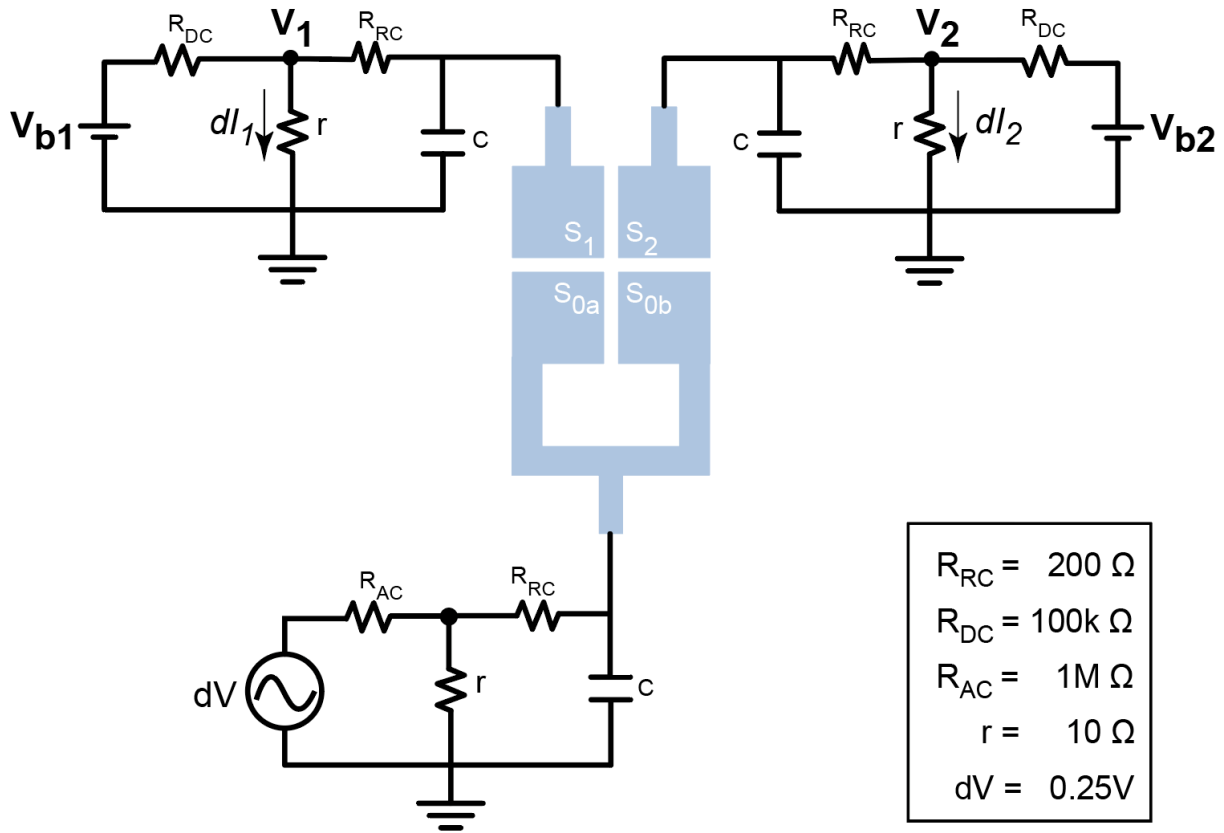

Fig. S1: Schematic diagram of the dual voltage source for quartet measurement.

## S2. Comparison of currents at $S_1$ and $S_2$

Figure S2 shows the conductance $G_i = dI_i/dV$ measured at terminals $S_1$ and $S_2$, respectively. Similar features can be found in both, including the Josephson currents between any two leads and the quartet current along $V_1 = -V_2$. Overall the conductance at $S_2$ is lower than that at $S_1$, suggesting the asymmetric couplings between each pair of contacts.

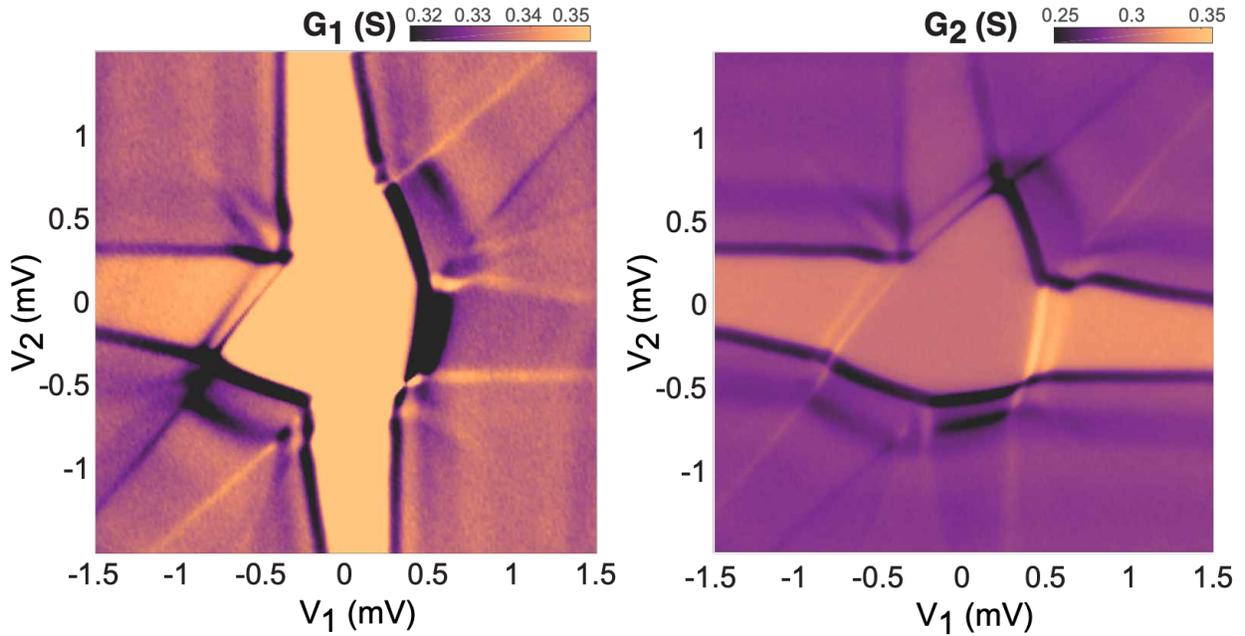

Fig. S2: Comparative data between $G_1 = dI_1/dV$ and $G_2 = dI_2/dV$ as a function of $V_1$ and $V_2$.

## S3. Quartet on top of the background quasiparticle signal

The quartet signal is much weaker compared to the Josephson signal. Besides the fact that quartet is an $8^{th}$-order process as opposed to the $4^{th}$-order of a Josephson current, the quartets exist outside of the zero-bias-voltage region, where signals from multiple Andreev reflections (MAR) are in the background. Figure S3a shows the color plot of the conductance $G_2 = dI_2/dV$ as a function of the bias $V_1$ and magnetic field while $V_2$ is fixed at 6 V. Figure S3b shows the zoom-in scan around the quartet signal. The oscillations in magnetic field show a 3% variation of the total conductance (Fig. S3c).

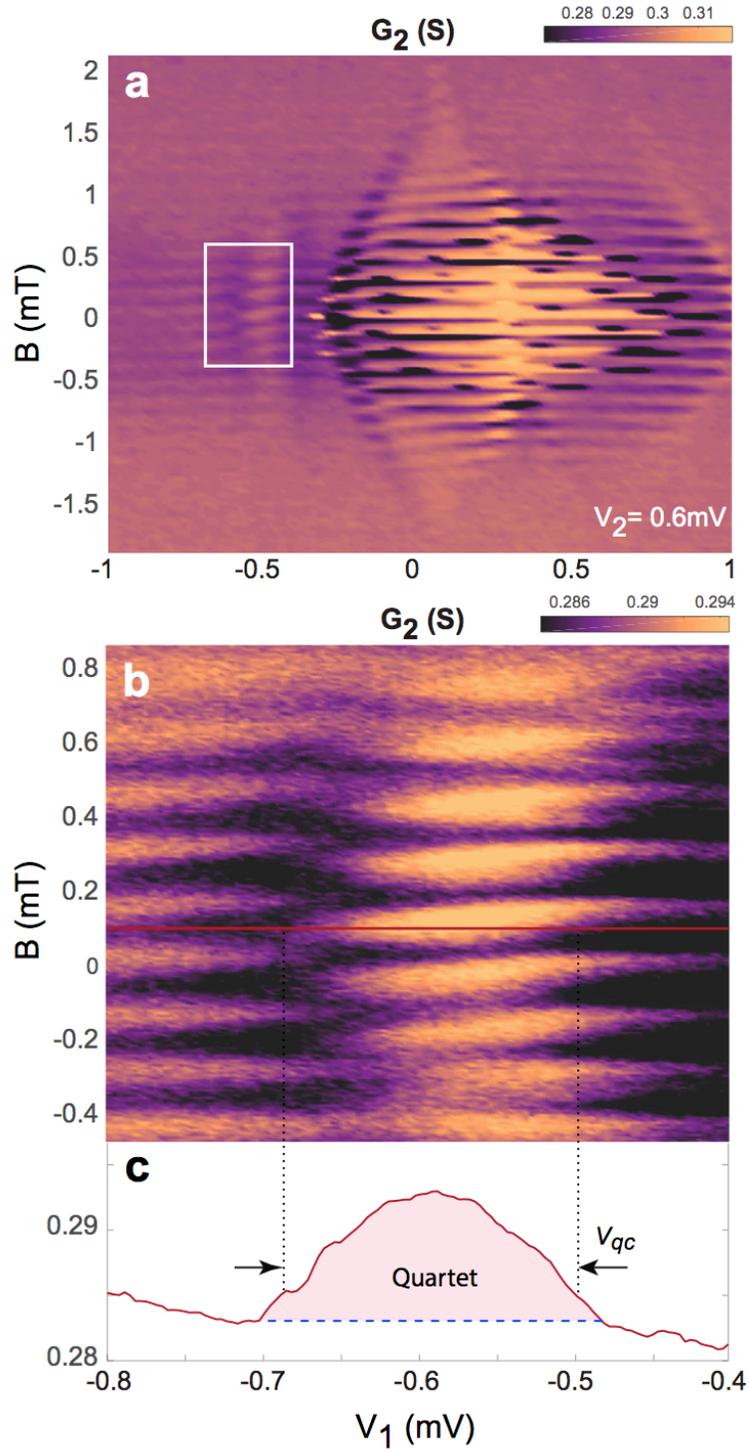

Fig. S3: **a**. $G_2 = dI_2/dV$ as a function of the DC bias voltage $V_1$ and magnetic field when $V_2$ is fixed at 0.6 V. The field-dependent quartet supercurrent is in the white box and the zoom-in scan is shown in **b**. Panel **c** shows the quartet signal on top of the quasiparticle background (blue dashed line). $V_{qc}$ is the voltage corresponding to the quartet critical current $I_{qc}$.

## S4. Gate dependence of the multi-terminal graphene Josephson junction

By applying a gate voltage, we can tune the chemical potential of the graphene channel region. Moreover, the gate voltage can change the density of states of the graphene underneath Al contacts, modulating the couplings between the graphene and the superconductors. Figure S4 shows the differential conductance $G_1$, measured at between $S_1$ and the grounded loop $S_0$, as a function of the two DC bias voltages $V_1$ and $V_2$ at back-gate $V_{bg}$= -10 V, -5 V, and 40 V. The critical value of each supercurrent is modulated accordingly, as well as the central zero-bias region. Note that among these back-gate voltages, the quartet supercurrent is the strongest at $V_{bg}$= -5 V.

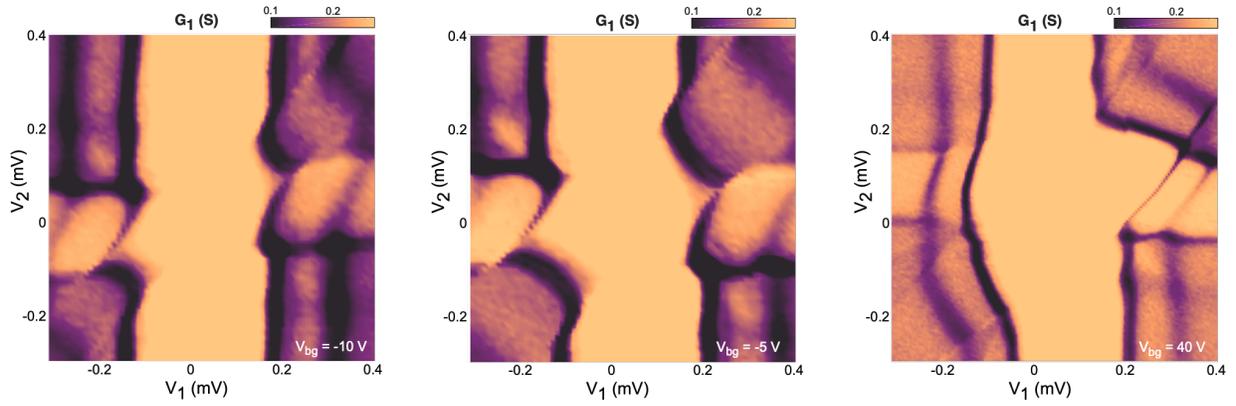

Fig. S4: $G_1(V_1, V_2)$ at $V_{bg}$ = -10 V, -5 V, and 40 V. Among these back-gate voltages, the quartet supercurrent (along $V_1 = -V_2$) is the strongest at $V_{bg}$= -5 V.

## S5. Theoretical complement for the interpretation of the experiment

This work focuses on the conductance modulation along the quartet line, $V_1 = -V_2 = V$, due to an applied flux. Along this line, where the microscopic models are solved, the quartet phase $\varphi_q$ is stationary while the other conjugated phase variable $\varphi_r$ is time dependent. For a general JJ at a finite temperature, zero-bias conductance increases monotonically as the critical current increases [1]. In our multi-terminal JJ, a similar scenario occurs along the quartet line, where the amplitude of the quartet conductance $G_1(V, -V)$ increases monotonically as the critical quartet current $I_{qc}(V, \Phi)$ increases [2]. Experimentally this is shown in Fig. S5. Therefore, the modulation of $G_1$ and $G_2$ as a function of the applied flux and on the applied quartet bias reflect the modulation of

$I_{qc}(V,\Phi)$. Models present in reference [3-5] provide the « quartet current »-« quartet phase » characteristics $I_q(\varphi_q, V, \Phi)$ and one takes $I_{qc}(V,\Phi) = Max[I_q(\varphi_q, V, \Phi)]$ on $\varphi_q$.

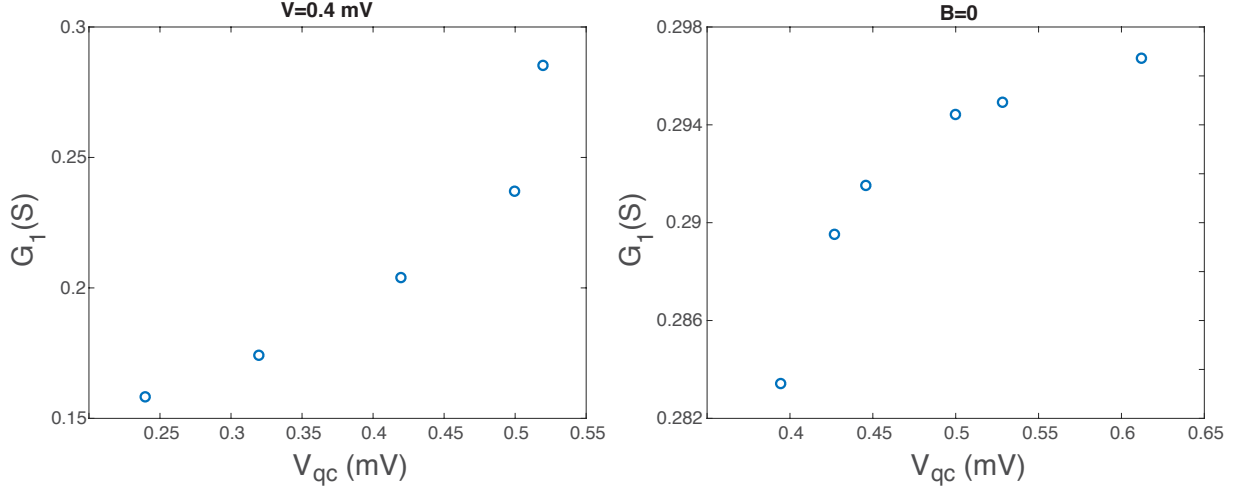

Fig.S5. Experimentally, the quartet conductance $G_1$ along the $V_1 = -V_2 = V$ indeed is found to increase monotonically as $V_{qc}$ increases at a fixed bias voltage with varying magnetic flux (left panel). Same monotonic relation between the quartet conductance and quartet critical current for a fixed magnetic field with varying bias voltage $V$ is shown in the right panel.

### A. Flux periodicity of the conductance

The present device contains a loop in the « quartet source terminal » $S_0$. The resulting interferences demonstrate the phase coherence of the DC quartet mode, despite the presence of biases. Moreover, they exhibit oscillations with period $\Phi_0/2 = hc/4e$ instead of $\Phi_0$. This is deeply different from a fundamental $\Phi_0$ and a first harmonic $\Phi_0/2$, that would happen for a transparent SQUID made of two ordinary two-terminal junctions: here, (i) the loop only encloses only one junction, (ii) the quartet current flows from both branches $S_{0a}$, $S_{0b}$ towards contacts $S_1$, $S_2$ and (iii) the voltage bias prohibits DC Cooper pair current. So the correct interpretation of those two periodicities is instead: the $\Phi_0/2$ period is the fundamental one, it signals the interference of quartets, revealing their charge $4e$. And the $\Phi_0$ period is a *subharmonic* due to the possibility of splitting the quartets. The flux indeed directly affects the quartet process and can be used as a probe of the nontrivial effect of the voltage. This is essential for the interpretation of the non-monotonous

$G(V)$ variation found in Fig. 4 (main text). Note that the « inversion » found in $I_c(\Phi)$ as $V$ increases, namely $I_{qc}(\Phi = 0) < I_{qc}(\Phi = \Phi_0/2)$ is counterintuitive: one would naively expect that destructive interferences in the $S_0$ loop decrease the quartet channel.

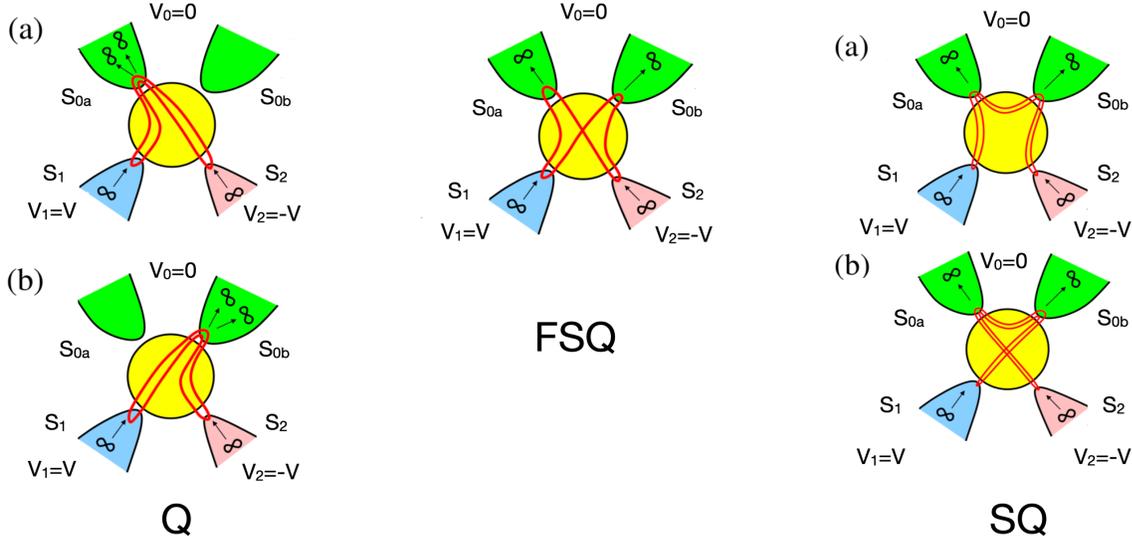

Fig.S6. Representation of the lowest order quartet processes. (Left) Q-processes. (Middle) FSQ-process, a split quartet process that vanishes by statistical fluctuations in a multichannel diffusive junction. (Right) SQ-processes.

The meaning of such an inversion is revealed by the perturbative diagrams shown on Fig. S6. The number of lines representing propagating amplitudes is limited by the transparency of the Graphene-Aluminum interfaces. First, the quartet « source » which is the grounded terminal can be either $S_{0a}$ or $S_{0b}$. In this case, both pairs forming the quartet emerge from the same branch $S_{0a}$ or $S_{0b}$, which we denote as Q processes. Alternatively, one pair can emerge from $S_{0a}$ and one from $S_{0b}$. The diagram in the center of Figure S6-1 corresponds to « splitting » the quartet in two entangled processes due to the exchange of two fermions [5]. We call this diagram FSQ (Fluctuating Split Quartet) because in a multichannel junction, due to spatial phase fluctuations, the dominant split quartet diagram that survives disorder has two more propagating lines (SQ) and involves the exchange of a quasiparticle between terminals $S_{0a}, S_{0b}$.

The two Q-processes differ by the phase acquired by four electrons instead of two for an usual superconducting loop, yielding a periodicity $\Phi_0/2 = hc/4e$, whatever the interface transparencies at the contacts. In terms of phase, going from Q diagrams to SQ (FSQ) diagrams formally transfers one pair of the quartet $S_{0a}$ (resp. $S_{0b}$) to $S_{0b}$ (resp. $S_{0a}$), which implies a phase change $2\pi\Phi/\Phi_0$. Therefore, the *interference* between processes Q and processes SQ (FSQ) creates a periodicity $\Phi_0$ that reminds the one of an ordinary SQUID.

Explicating the quartet phase, one can write the sum of processes Q, SQ (or FSQ) as:
$$I(\varphi_q, \Phi) = I_{c,Qa}\sin(\varphi_q - \Phi) + I_{c,Qb}\sin(\varphi_q + \Phi) + I_{SQ}\sin\varphi_q$$
Here the prefactors can take positive or negative values. Quartets are generically $\pi$-junctions in the tunnel regime and at vanishing voltage, i.e., $I_{c,Qi} < 0$, (i=a, b). Taking the maximum with respect to $\varphi_q$ of $I(\phi_q, \Phi)$ gives the critical current. For $\Phi = 0$ one has $I_{qc}(0) = |I_{c,Qa} + I_{c,Qb} + I_{SQ}|$ while for $\Phi = \Phi_0/2$ one has $I_{qc}(\Phi_0/2) = |I_{c,Qa} + I_{c,Qb} - I_{SQ}|$. The relative sign of $I_{c,Qa,b}$ and $I_{SQ}$ thus governs the flux variation of $I_{qc}$. If it is positive, the total quartet critical current has a maximum at $\Phi = 0$ and a minimum at $\Phi = \Phi_0/2$, ie $I_{qc}(\Phi = 0) > I_{qc}(\Phi = \Phi_0/2)$. If it is negative, the minima and maxima are inverted and $I_{qc}(\Phi = 0) < I_{qc}(\Phi = \Phi_0/2)$. Importantly, a pure $\Phi_0/2$ periodicity implies that the SQ channel is absent, leaving only the Q contributions. In a perturbative limit, no inversion is found. The next section explains how the new control variable $V$ can indeed trigger an inversion of the sign of $I_{qc}(\Phi = 0) - I_{qc}(\Phi = \Phi_0/2)$.

### B. The dot model, numerical results and the Floquet-Landau-Zener interpretation
#### a. The dot model.
The dot model [3] treats the metallic junction as a 0D object with a single level having one-electron energy $\epsilon_0/2$, no Coulomb interaction and coupled to the four terminals by one-electron matrix elements $t_i e^{i\phi_i/2}$. Here i=(1, 2, 0a, 0b) and the superconducting phases are incorporated in the matrix elements. With a convenient gauge choice, the phases $\varphi_i$ are
$\varphi_1 = [\varphi_q + \varphi_r(t)]/2, \varphi_2 = [\varphi_q - \varphi_r(t)]/2, \varphi_{0a} = \Phi/2, \varphi_{0b} = -\Phi/2$ with $\varphi_r(t) = 2eVt/\hbar$.

The Hamiltonian writes:

$$H_{dot}(t) = \sum_{ik\sigma} \epsilon_{ik\sigma} c^\dagger_{ik\sigma} c_{ik\sigma} + \sum_{ik} \Delta(c^\dagger_{ik\uparrow} c^\dagger_{i,-k\downarrow} + H.c.) + \sum_{ik\sigma} [t_i e^{i\phi_i(t)/2} c^\dagger_{ik\sigma} d_{i\sigma} + H.c.]$$
$$+ \epsilon_0 \sum_\sigma d^\dagger_\sigma d_\sigma$$

The dot energy $\epsilon_0$ mimics the control of the metallic junction by a gate. This model can be generalized to several noninteracting levels with energies $\epsilon_i$.

The model can be further simplified in the large gap limit $\Delta \gg \epsilon_0, t_i$. Single particle processes are eliminated out and only pair processes remain:
$$H_{dot,\infty}(t) = \epsilon_0(2b^\dagger b - 1) + \Gamma(t) b + \Gamma^*(t) b^\dagger$$
with $b = d_\downarrow d_\uparrow$ and $\Gamma(t) = \Gamma_{0a} e^{i\Phi/2} + \Gamma_{0b} e^{-i\Phi/2} + \Gamma_1 e^{i[\varphi_q+\varphi_r(t)]/2} + \Gamma_2 e^{i[\varphi_q-\varphi_r(t)]/2}$
with $\Gamma_i \sim \pi t_i^2 N(0)$ where $N(0)$ is the Aluminum density of states. The reduced Hamiltonian $H_{dot,\infty}(t)$ describes a driven two-level system and it can be solved with Floquet techniques. In fact the Hamiltonians $H_{dot}(t)$ and $H_{dot,\infty}(t)$ describe a system periodically driven by the running phase $\phi_r(t) = 2eVt/\hbar$. This can be compared to the driving of a Josephson junction by a microwave field, with an important difference: here the drive amplitude is given by $\Gamma_1, \Gamma_2$ and is *non-perturbative*. Yet, there are similarities: an adiabatic regime holds when the drive frequency is much smaller than the equilibrium Andreev gap. And when it is comparable, Landau-Zener-like transitions occur that couple non-perturbatively the two Andreev levels [6].

Let us comment on the effect of the flux in the large gap limit. $\Gamma(t)$ can be rewritten as
$$\Gamma(t) = \Gamma_0(\Phi) e^{i\alpha(\Phi)} + \Gamma_1 e^{i[\varphi_q+\varphi_r(t)]/2} + \Gamma_2 e^{i[\varphi_q-\varphi_r(t)]/2}$$
with $\Gamma_0(\Phi) = \sqrt{\Gamma_{0a}^2 + \Gamma_{0b}^2 + 2\Gamma_{0a}\Gamma_{0b}\cos\Phi}$, $\tan\alpha(\Phi) = \frac{\Gamma_{0a}-\Gamma_{0b}}{\Gamma_{0a}+\Gamma_{0b}} \tan\Phi/2$

Therefore, in this limit the role of the loop is to map the four-terminal junction onto a three-terminal one with a *flux-dependent* coupling to S_c, together with a phase shift $\alpha(\Phi)$ that can be absorbed in the definition of the quartet phase $\varphi_q$. Nevertheless, the dependence of the quartet dynamics on the coupling $\Gamma_c$ is far from trivial, it is non-monotonous and thus the flux knob reveals the non-adiabatic effects that explain the experimental observation.

**b. The adiabatic solution.**

At a given time t, the instantaneous Andreev Bound State (ABS) energies are given by

$E_{ABS}(t) = \pm\sqrt{\epsilon_0^2 + |\Gamma(t)|^2}$. In the limit $eV \ll (\epsilon_0, \Gamma_i) \ll \Delta$, one can average out the slow drift motion of the phase. $\varphi_r(t)$ i.e. on the period $T = h/2eV$ to yield adiabatic ABS, that only depend on the quartet phase $\varphi_q$ and on the flux $\Phi$.

$$\langle E_{ABS}\rangle(\varphi_q, \Phi) = \pm\frac{1}{T}\int_0^T \sqrt{\epsilon_0^2 + |\Gamma(t)|^2}\,dt$$

Importantly, these effective adiabatic ABS do not depend on the voltage $V$. They give rise to an adiabatic quartet current $I_{q,adiab} = (2e/\hbar)\partial\langle E_{ABS}\rangle/\partial\varphi_q$.

In the related case of a microwave-irradiated junction with frequency $\omega$, the adiabatic approximation is controlled both by the smallness of the parameter $\hbar\omega/\delta_{ABS}$ (where $\delta_{ABS}$ is the ABS mini gap at equilibrium) and by the small amplitude of the microwave field. Remind that in our case, even if $\delta_{ABS}$ is not zero and $eV/\delta_{ABS}$ is small, the periodic drive is strong.

### d. Numerical solution of the dot model.

As $V$ increases in the experiment, the conductance oscillation has main period $\Phi_0$, then its frequency doubles with period $\Phi_0/2$, and at higher $V$ it recovers the period $\Phi_0$ but with a phase shift equal to $\pi$. The frequency doubling corresponds to vanishing of SQ (FSQ) processes, leaving nearly pure quartet $Q_a$, $Q_b$ processes interfering. Therefore, at a crossover voltage there is a change of the relative signs of the Q and SQ (FSQ) components.

This scenario is supported by numerical results obtained at nonzero voltage with the dot model and a finite superconducting gap. We use Keldysh non-equilibrium Green's functions to solve the dynamics of the model described by Hamiltonian $H_{dot}(t)$ [3]. The quartet current-phase characteristics is obtained for any $V$ and $\Phi$. Typical parameters are $\epsilon_0 = 0$ (resonant dot) and:

$$\Gamma_1 = 0.4\Delta, \Gamma_2 = 0.2\Delta, \Gamma_{0a} = (0.3\Delta + \gamma)/2, \Gamma_{0b} = (0.9\Delta + \gamma)/2$$

where $\gamma$ controls the relative strength of the $S_0$ coupling relatively to the $S_1, S_2$ couplings. More generally, to obtain a $\pi$-shift of the $I_{qc}(\Phi)$ oscillation one needs: (i) a minimum asymmetry $\frac{|\Gamma_{0a}-\Gamma_{0b}|}{\Gamma_{0a}+\Gamma_{0b}}$ of the contacts $S_{0a}, S_{0b}$, (ii) a minimum ratio $\frac{\Gamma_{0a}+\Gamma_{0b}}{\Gamma_1+\Gamma_2}$, (iii) small enough $\epsilon_0$.

Figure S7 shows the quartet critical current as a function of $V$, for both $\Phi = 0$ and $\Phi = \Phi_0/2$ (with $\gamma = 0.3$). It contains two inversion regions, the main one being at higher voltage. Other narrower inversions are also observed at lower voltages (not represented here).

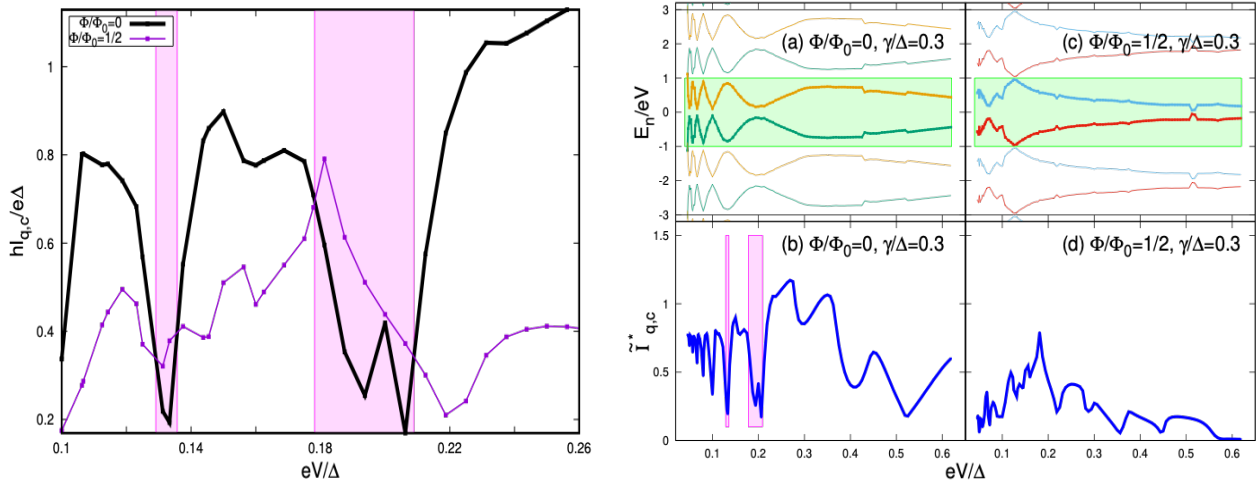

Figure S7. (Left) Maximum quartet current for $\Phi = 0$ (black) and $\Phi = \Phi_0/2$ (red), showing two « inversion » windows. (Right, bottom) Same but on an extended voltage range. (Right, top) The corresponding Andreev-Floquet ladders, showing the correlation of the anticrossings with the minima of $I_{qc}$.

**e. The Floquet ladders and the Landau-Zener resonance.**

In this model, the inversion relies on the periodic drive forming Floquet bands from the initial equilibrium ABS. The adiabatic energies $\langle E_{ABS} \rangle$ allow to form Floquet ladders. Here the quartet phase plays the same role as a one-dimensional crystal momentum and the ladders are similar to Wannier-Stark ladders [4]. The Floquet ladders replicate $\langle E_{ABS} \rangle$, following the « classical » formula:

$$E_{n,\pm,class}(\phi_q, \Phi, V) = \pm \langle E_{ABS} \rangle (\phi_q, \Phi) + 2neV$$

This picture is correct as far as the Floquet bands are far from each other, i.e. when their spacing $2eV$ is much larger than their dispersion $\sim \Gamma$. Decreasing $V$, when two Floquet bands $E_{n,+,class}$ and $E_{m,-,class}$ cross each other, an avoided crossing happens. This correspond to a resonance of first order if $|n - m| = 1$. Due to the non-perturbative character of the periodic drive, each Floquet band is affected by the presence of its neighbors or of others (higher-order resonances). Anticrossings are due to Landau-Zener-Stückelberg transitions in the instantaneous ABS spectrum

$E_{ABS}(t)$: as the running phase slowly drifts, non-adiabatic transitions are obtained close to the minima of the ABS gap $\sqrt{\epsilon_0^2 + |\Gamma(t)|^2}$, which can even vanish if $\epsilon_0 = 0$.

Anticrossings and deformations of the Floquet bands manifest a nonlinear dependence of the spectrum $E_{n,\pm}(\varphi_q, \Phi, V)$ with the voltage. Let us now relate the inversions of $I_{qc}(\Phi)$ to anticrossings of « classical » Floquet bands.

Figure S7 also shows a part of the Floquet spectrum (inside the superconducting gap) as a function of the reduced voltage $eV/\Delta$, in a more extended voltage range. The plot represents $E_{n,\pm}/eV$. The anticrossings are visible around precise voltage values, and they get more and more frequent at low voltage. The quartet critical current $I_{qc}$ is represented for both $\Phi = 0$ and $\Phi = \Phi_0/2$. The minima of $I_{qc}$ are perfectly correlated to the resonances. Like for a microwave-irradiated junction, the anticrossings manifest a quantum mixing of the adiabatic states $(n+, m-)$, that bear opposite indices therefore carry opposite currents. As a consequence, in the region of the anticrossing, quantum fluctuations reduce the quartet current $I_{qc}$.

Let us now consider the flux dependence of the position of these resonances and $I_{qc}$ minima. Roughly speaking, the role of the flux is to modulate the coupling $\Gamma_c(\Phi)$. This in turn modifies the ABS spectrum, its gap and the rate of Landau-Zener transitions. The position of the resonances on the V-axis therefore oscillate periodically with the flux. Figure S7 indeed shows that the minima of $I_{qc}(\Phi = 0)$ and those of $I_{qc}(\Phi = \Phi_0/2)$ do not coincide. This causes a crossing of the curves representing $I_{qc}(V, \Phi = 0)$ and $I_{qc}(V, \Phi = \Phi_0/2)$. The phenomenon is quite robust against a variation of the couplings $\Gamma_i$.

One sees on Figure S8 that below and above the resonance (top and bottom panels), the quartet characteristics is a $\pi$-junction, with essentially the period $\Phi_0$. Close to the inversion point (central panels) it has a strong $\Phi_0/2$ period component featuring Q processes.

Let us comment on the validity of the single level dot model. A numerical study of a many-level dot model was also carried out, in a wide range of parameters so as to mimic a multichannel junction. Generically, regions of inversion are also found as the voltage is varied, showing the robustness of the phenomenon. Finally, preliminary calculations indicate that in a more general picture of a metallic junction, a more general reasoning based on Keldysh Green's function also gives rise to inversions. More theory is found in forthcoming papers [5].

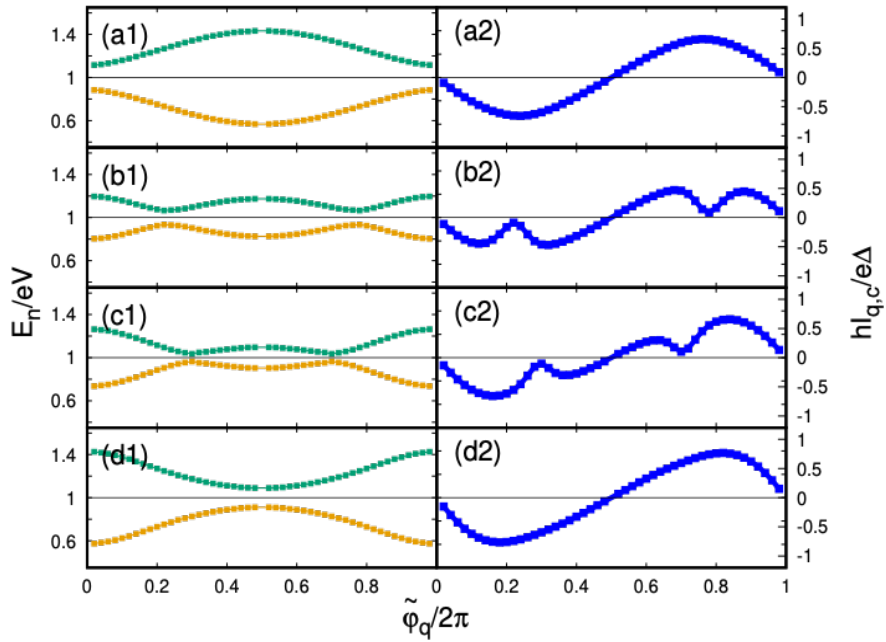

Figure S8. (Right) Evolution of the quartet current-phase characteristics through a Landau-Zener resonance (from a2 to d2). (Left) The corresponding evolution of the Andreev-Floquet bands.